\documentclass[aps,pre,superscriptaddress,twocolumn,longbibliography,floatfix]{revtex4-2}

\usepackage{graphicx}
\usepackage{dcolumn}
\usepackage{bm}
\usepackage{algorithm}
\usepackage{algpseudocode}
\usepackage[colorlinks]{hyperref}
\usepackage{amsmath,amsthm,amssymb}
\usepackage{mathrsfs}
\usepackage{multirow}
\usepackage[all]{xy}
\usepackage{pbox}
\usepackage{verbatim}
\usepackage{braket}
\usepackage{mathtools}
\usepackage{tikz}
\usepackage{xcolor}
\usepackage{xfrac}
\usepackage{cleveref}

\newcommand{\singlefigure}{0.45\textwidth}

\begin{document}
\title{Physics-Informed Long-Range Coulomb Correction for Machine-learning Hamiltonians}
\author{Yang Zhong}
\email{yzhong@fudan.edu.cn}
\affiliation{Key Laboratory of Computational Physical Sciences (Ministry of Education), Institute of Computational Physical Sciences, State Key Laboratory of Surface Physics, and Department of Physics, Fudan University, Shanghai, 200433, China.}
\author{Xiwen Li}
\affiliation{Key Laboratory of Computational Physical Sciences (Ministry of Education), Institute of Computational Physical Sciences, State Key Laboratory of Surface Physics, and Department of Physics, Fudan University, Shanghai, 200433, China.}
\author{Xingao Gong}
\affiliation{Key Laboratory of Computational Physical Sciences (Ministry of Education), Institute of Computational Physical Sciences, State Key Laboratory of Surface Physics, and Department of Physics, Fudan University, Shanghai, 200433, China.}
\author{Hongjun Xiang}
\email{hxiang@fudan.edu.cn}
\affiliation{Key Laboratory of Computational Physical Sciences (Ministry of Education), Institute of Computational Physical Sciences, State Key Laboratory of Surface Physics, and Department of Physics, Fudan University, Shanghai, 200433, China.}

\begin{abstract}
  Machine-learning electronic Hamiltonians achieve orders-of-magnitude speedups over density-functional theory, yet current models omit long-range Coulomb interactions that govern physics in polar crystals and heterostructures. We derive closed-form long-range Hamiltonian matrix elements in a nonorthogonal atomic-orbital basis through variational decomposition of the electrostatic energy, deriving a variationally consistent mapping from the electron density matrix to effective atomic charges. We implement this framework in HamGNN-LR, a dual-channel architecture combining E(3)-equivariant message passing with reciprocal-space Ewald summation. Benchmarks demonstrate that physics-based long-range corrections are essential: purely data-driven attention mechanisms fail to capture macroscopic electrostatic potentials. Benchmarks on polar ZnO slabs, CdSe/ZnS heterostructures, and GaN/AlN superlattices show two- to threefold error reductions and robust transferability to systems far beyond training sizes, eliminating the characteristic staircase artifacts that plague short-range models in the presence of built-in electric fields.
\end{abstract}
\maketitle
\section{Introduction}
Accurate electronic-structure calculations underpin condensed-matter physics and materials science. Despite the success of density functional theory (DFT)~\cite{jones2015density,becke2014perspective}, self-consistent field (SCF) iterations entail prohibitive computational cost for large or complex systems. Machine-learning electronic Hamiltonians offer a transformative alternative~\cite{schutt2019unifying,unke2021se,li2022deep,gong2023general,zhong2023transferable,gu2024deeptb}: by mapping atomic configurations directly to Hamiltonian matrix elements, these models bypass the SCF loop and achieve orders-of-magnitude speedups while preserving first-principles accuracy. Yet prevailing approaches invoke the nearsightedness principle~\cite{prodan2005nearsightedness}, confining information flow to local atomic neighborhoods. This approximation breaks down in systems governed by long-range Coulomb interactions, including polar crystals, low-dimensional materials, and heterostructures, in which charge transfer, built-in electric fields, and macroscopic polarization arise from electrostatic interactions that extend far beyond any local cutoff.

In the development of machine-learning interatomic potentials, long-range interactions have been incorporated through charge equilibration~\cite{ko2021fourth}, Gaussian charge representations~\cite{zhang2022deep}, Ewald summation for $1/r$ Coulomb tails~\cite{kosmala2023ewald}, Fourier convolutions~\cite{grisafi2019incorporating}, and reciprocal-space neural networks~\cite{yu2022capturing}. These techniques target scalar potential energy surfaces and do not extend to electronic Hamiltonian matrices, which are tensorial matrix quantities constrained by E(3) equivariance and time-reversal symmetry. The mathematical structure required to encode long-range physics in Hamiltonians thus differs fundamentally from that in potential models.

We present a rigorous framework for incorporating long-range Coulomb interactions into Machine-learning electronic Hamiltonians in a nonorthogonal atomic-orbital (NAO) basis. Through variational decomposition of the electrostatic energy into short- and long-range contributions, we derive closed-form expressions for long-range Hamiltonian matrix elements that ensure exact consistency between the Hamiltonian and total energy. This reciprocal-space formulation enables efficient evaluation via rapidly convergent $k$-space sums and systematic treatment of polar crystals and heterostructures.

\section{Results}
We decompose the total energy as $E_{\text{tot}} = E_{\text{sr}} + E_{\text{lr}}$, isolating the slowly convergent Coulomb tail into a reciprocal-space contribution~\cite{ewald1921calculation}. With cutoff wave vector $k_c$ and Gaussian smoothing width $\sigma$, the long-range electrostatic energy is
\begin{equation}
  E_{\text{lr}} = \frac{1}{2\epsilon_0 V} \sum_{0 < |\mathbf{k}| < k_c} \frac{1}{k^2} \exp(-\sigma^2 k^2 / 2) \left| S(\mathbf{k}) \right|^2,
  \label{eq:longrange-energy}
\end{equation}
where $V$ is the unit-cell volume, $\epsilon_0$ the vacuum permittivity, and $S(\mathbf{k}) = \sum_i Q_i \exp(i \mathbf{k} \cdot \bm{\tau}_i)$ is the structure factor of net atomic charges $Q_i$ at site $i$. To evaluate Eq.~(\ref{eq:longrange-energy}), we require the net atomic charges $Q_i = Z_i - q_i$, where $q_i$ is the electronic population on atom $i$. We obtain $q_i$ by integrating the electron density $\rho(\mathbf{r})$ against a spatially localized weight function centered on atom $i$, following density-based population analysis~\cite{hirshfeld1977,becke1988multicenter}:
\begin{equation*}
  q_i = \int W_i(\mathbf{r} - \bm{\tau}_i)\,\rho(\mathbf{r})\,\mathrm{d}\mathbf{r}.
\end{equation*}
In a nonorthogonal atomic-orbital (NAO) basis, $\rho(\mathbf{r}) = \sum_{\mu\nu} P_{\mu\nu}\,\phi_\mu^*(\mathbf{r})\,\phi_\nu(\mathbf{r})$, so $q_i = \sum_{\mu\nu} P_{\mu\nu}\,W_{i,\mu\nu}$ with orbital-projected weight-matrix elements
\begin{equation*}
  W_{i,\mu\nu} \;\equiv\; \int W_i(\mathbf{r} - \bm{\tau}_i)\,\phi_\mu^*(\mathbf{r})\,\phi_\nu(\mathbf{r})\,\mathrm{d}\mathbf{r} \;=\; \frac{\partial q_i}{\partial P_{\mu\nu}}.
\end{equation*}
Since $Z_i$ is constant, we have $\partial Q_i/\partial P_{\mu\nu} = -W_{i,\mu\nu}$. The long-range Hamiltonian matrix elements follow from the chain rule:
\begin{equation}
  H_{\text{lr},\mu\nu}
  \;=\;\frac{\partial E_{\text{lr}}}{\partial P_{\mu\nu}}
  \;=\;\sum_i \frac{\partial E_{\text{lr}}}{\partial Q_i}\,\frac{\partial Q_i}{\partial P_{\mu\nu}},
  \label{eq:chainrule}
\end{equation}
where $\partial E_{\text{lr}}/\partial Q_i$ is obtained by differentiating $|S(\mathbf{k})|^2$; since $\partial S/\partial Q_i = \exp(i\mathbf{k}\cdot\bm{\tau}_i)$, the cross terms yield $2\operatorname{Re}[S(\mathbf{k})\,\exp(-i\mathbf{k}\cdot\bm{\tau}_i)]$ (Appendix~\ref{app:derivation}). For a real-valued NAO basis, always available when time-reversal symmetry holds, $W_{i,\mu\nu}$ is real. Substituting into Eq.~(\ref{eq:chainrule}) and converting to atomic units ($4\pi\epsilon_0 = 1$) yields the closed-form result:
\begin{equation}
  \begin{split}
    H_{\text{lr},\mu\nu} = -\frac{4\pi}{V} \sum_{0 < |\mathbf{k}| < k_c} \frac{\exp(-\sigma^2 k^2 / 2)}{k^2} \\
    \times \operatorname{Re}\left[ S(\mathbf{k}) \sum_i \exp(-i \mathbf{k} \cdot \bm{\tau}_i) W_{i,\mu\nu} \right].
  \end{split}
  \label{eq:longrange-ham}
\end{equation}
Equation~(\ref{eq:longrange-ham}) couples the macroscopic field
carried by $S(\mathbf{k})$ to the orbital-projected weight matrix
$W_{i,\mu\nu}$, ensuring variational
consistency---$H_{\text{lr},\mu\nu}
  = \partial E_{\text{lr}}/\partial P_{\mu\nu}$---by construction.
The Ewald self-interaction energy
$E_{\text{self}} \propto \sum_i
  Q_i^2$~\cite{toukmaji1996ewald,deleeuw1980ewald} generates a
spatially local correction
$\partial E_{\text{self}}/\partial P_{\mu\nu} \propto \sum_i
  Q_i\,W_{i,\mu\nu}$ that lies within the short-range receptive
field and is absorbed into $H_{\text{sr}}$
(Appendix~\ref{app:self}). Expanding $S(\mathbf{k}) = \sum_j Q_j\,e^{i\mathbf{k}\cdot
  \bm{\tau}_j}$ shows that the bracketed factor in
Eq.~(\ref{eq:longrange-ham}) reduces to
$\operatorname{Re}\bigl[\sum_{i,j}
  Q_j\,W_{i,\mu\nu}\,e^{-i\mathbf{k}\cdot(\bm{\tau}_i -
  \bm{\tau}_j)}\bigr]$, a pairwise sum whose phases depend solely
on relative displacements $\bm{\tau}_i - \bm{\tau}_j$. This
structure is directly exploited by the Ewald attention module
described below.

\begin{figure}
  \centering
  \includegraphics[width=\singlefigure]{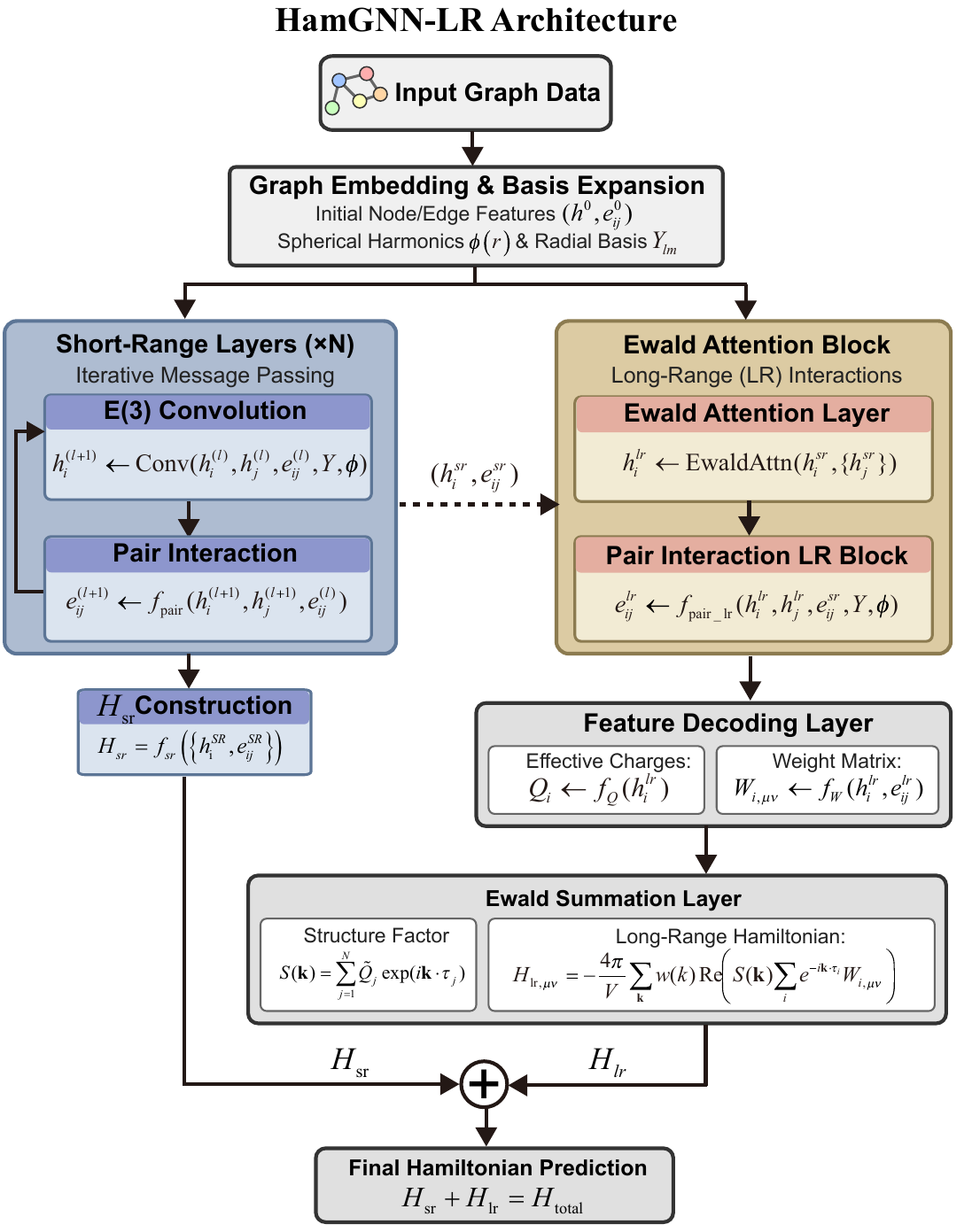}
  \caption{\label{fig1} Architecture of HamGNN-LR for predicting $H_{\text{total}} = H_{\text{sr}} + H_{\text{lr}}$. Short-range E(3)-equivariant message passing constructs $H_{\text{sr}}$ from local environments. An Ewald attention module captures long-range correlations in reciprocal space with $\mathcal{O}(N|\mathcal{K}|)$ cost, linear in system size $N$ for fixed sampled wave vectors $|\mathcal{K}|$, and decodes ionic charges $Q_i$ and weight matrices $W_{i,\mu\nu}$. The decoded charges are neutralized ($Q_j \!\mapsto\! \tilde{Q}_j$) and, together with the weight matrices $W_{i,\mu\nu}$, substituted into the analytic correction $H_{\text{lr}}$ [Eq.~(\ref{eq:longrange-ham})].}
\end{figure}

Guided by this analytic form, we construct HamGNN-LR
(Fig.~\ref{fig1}), a dual-channel architecture separating
short- and long-range physics. The short-range channel employs
E(3)-equivariant message passing for local bonding and orbital
hybridization~\cite{zhong2023transferable,zhong2026unihamgnn},
producing equivariant node features $\mathbf{h}_i^{\text{sr}}$.
The long-range channel---an \emph{Ewald attention
  module}---operates in reciprocal space to capture macroscopic
electrostatic correlations beyond the real-space cutoff, and
ultimately evaluates $H_{\text{lr}}$ via
Eq.~(\ref{eq:longrange-ham}). The two channels are merged
through a gated residual connection
$\mathbf{h}_i^{\text{lr}} \leftarrow
  (1-\alpha)\,\mathbf{h}_i^{\text{sr}}
  + \alpha\,\mathrm{EquivariantNonlin}(\mathbf{h}_i^{\text{lr}})$
with learnable gate $\alpha$, ensuring that the long-range
correction is introduced as a small perturbation to the
short-range baseline during early training.

The Ewald attention module is designed to reproduce the
pairwise phase structure
$\mathbf{k}_m\!\cdot\!(\bm{\tau}_i - \bm{\tau}_j)$ that
appears in Eq.~(\ref{eq:longrange-ham}). We sample a discrete
set of wave vectors $\{\mathbf{k}_m\}$ with
$|\mathbf{k}_m| < k_c$ from the reciprocal lattice. For each
$\mathbf{k}_m$, invariant scalar components of the node
features are projected into query and key vectors, to which
rotary position encoding
(RoPE)~\cite{su2024roformer} is applied with phase
$\theta_{j,m} = \mathbf{k}_m\!\cdot\!\bm{\tau}_j$. The
resulting inner product
$\langle\tilde{\mathbf{q}}_{i,m},\,
  \tilde{\mathbf{k}}_{j,m}\rangle$ depends solely on the
relative displacement phase
$\mathbf{k}_m\!\cdot\!(\bm{\tau}_i - \bm{\tau}_j)$,
exactly matching the Ewald summation kernel. The value vectors
carry full equivariant content and are aggregated via
linearized attention---first summing over atoms at each
$\mathbf{k}_m$, then over wave vectors---yielding
$\mathcal{O}(N|\mathcal{K}|)$ complexity, linear in system
size $N$ for a fixed reciprocal grid, versus
$\mathcal{O}(N^2)$ for standard all-pairs attention. The aggregated
features $\mathbf{h}_i^{\text{lr}}$ inherit E(3) equivariance
because only invariant scalars participate in RoPE and
attention weights, while the equivariant value vectors pass
through unrotated.

From the updated long-range node features
$\mathbf{h}_i^{\text{lr}}$, the module decodes effective ionic
charges $Q_i$ via an invariant readout. Since
network-predicted charges need not satisfy global neutrality,
we apply a zero-mean projection
$\tilde{Q}_j = Q_j - N^{-1}\sum_k Q_k$ before Ewald
summation, defining the structure factor
\begin{equation}
  S(\mathbf{k}_m) = \frac{1}{N} \sum_{j=1}^N \tilde{Q}_j
  \exp(i\,\mathbf{k}_m \cdot \bm{\tau}_j),
  \label{eq:structure-factor-main}
\end{equation}
where the $1/N$ prefactor renders $S(\mathbf{k}_m)$ an
intensive quantity consistent with the $1/V$ normalization in
Eq.~(\ref{eq:longrange-energy}). Long-range edge features
$\mathbf{e}_{ij}^{\text{lr}}$ are constructed by concatenating
node features of the two endpoints and coupling them to
spherical-harmonic embeddings of the bond direction via
Clebsch--Gordan tensor products, injecting directional
information while preserving equivariance. The weight matrices
$W_{i,\mu\nu}$ are then decoded from the node features
(onsite elements) and edge features (offsite elements).
Finally, $\tilde{Q}_j$ and $W_{i,\mu\nu}$ are substituted
into Eq.~(\ref{eq:longrange-ham}) to yield $H_{\text{lr}}$,
giving the total prediction
$H_{\mu\nu} = H_{\text{sr},\mu\nu} + H_{\text{lr},\mu\nu}$.

Unlike prior reciprocal-space attention
schemes~\cite{kosmala2023ewald,frank2024efa,qu2026allscaip,
  caruso2025range}, which target scalar energy predictions, this
construction is grounded in the analytic Hamiltonian expression
  [Eq.~(\ref{eq:longrange-ham})], preserves $\mathrm{E}(3)$
equivariance at every stage, and scales linearly in $N$.
Additional implementation details---including Q/K/V
projections, the RoPE convention, linearized aggregation, and
the gated residual connection---are given in
Appendix~\ref{app:ewald_attention}.

To disentangle the effect of the explicit long-range correction
from that of increased network capacity, we restrict all models to a single
interaction layer, confining the receptive field to minimal
local neighborhoods. We compare four variants that
systematically include or exclude two key ingredients: the physics-based
$H_{\text{lr}}$ [Eq.~(\ref{eq:longrange-ham})] and the Ewald
attention module:
(i)~\textbf{SR}, a short-range baseline with neither;
(ii)~\textbf{LR-loc}, which evaluates $H_{\text{lr}}$ via
Eq.~(\ref{eq:longrange-ham}) but decodes $Q_i$ and
$W_{i,\mu\nu}$ from the short-range node and edge features
alone, without Ewald attention;
(iii)~\textbf{LR-EA} (full HamGNN-LR), which first extracts
long-range features through Ewald attention, decodes $Q_i$ and
$W_{i,\mu\nu}$ from the resulting long-range node and edge
features, and evaluates $H_{\text{lr}}$;
and (iv)~\textbf{EA}, which incorporates Ewald attention to
augment node features but omits the physics-based
$H_{\text{lr}}$ term, relying entirely on the network to learn
long-range effects in a data-driven manner.
Table~\ref{tab:layer1_comparison} reports Hamiltonian mean
absolute error (MAE) on three polar/heterostructure
systems---ZnO slab, GaN/AlN superlattice, and CdSe/ZnS
heterostructure---each split 60\%/20\%/20\% for
training/validation/test.

Incorporating the physics-based correction proves decisive. In
CdSe/ZnS, LR-EA reduces test MAE from 3.279~meV (SR) to
1.058~meV---a threefold improvement---while even LR-loc, which
lacks reciprocal-space feature extraction, reaches 1.292~meV;
in GaN/AlN, LR-EA similarly reduces test MAE from 3.646~meV to
1.162~meV. By contrast, EA---equipped with Ewald attention but
lacking the analytic Hamiltonian term---yields 3.158~meV in
CdSe/ZnS and 3.598~meV in GaN/AlN, essentially identical to
the short-range baseline. In all three tested systems, EA
provides negligible improvement, confirming that
data-driven attention alone, without the physics-based
Hamiltonian correction, is insufficient to capture macroscopic
electrostatic effects within a single-layer architecture. Both LR-loc and
LR-EA reduce errors by factors of two to three, demonstrating
that the explicit, variationally consistent $H_{\text{lr}}$ is
the essential ingredient; Ewald attention further enhances
accuracy by providing long-range-aware features for decoding
$Q_i$ and $W_{i,\mu\nu}$. Nevertheless, the truncated Coulomb
tail is recovered by the physics-based correction rather than by
the attention mechanism per se.

\begin{table}[htbp]
  \caption{\label{tab:layer1_comparison} Hamiltonian MAE (meV) for single-layer models on three polar/heterostructure systems. SR: short-range baseline; LR-loc: long-range correction [Eq.~(\ref{eq:longrange-ham})] with local decoding; LR-EA: full HamGNN-LR (correction + Ewald attention); EA: Ewald attention only, without physics-based $H_{\text{lr}}$.}
  \begin{ruledtabular}
    \begin{tabular}{llcccc}
      System                    & Split & SR    & LR-loc & LR-EA & EA    \\
      \hline
      \multirow{2}{*}{ZnO slab} & Train & 1.278 & 0.685  & 0.596 & 1.099 \\
                                & Test  & 1.336 & 0.747  & 0.635 & 1.153 \\
      \hline
      \multirow{2}{*}{GaN/AlN}  & Train & 3.623 & 1.462  & 1.184 & 3.556 \\
                                & Test  & 3.646 & 1.313  & 1.162 & 3.598 \\
      \hline
      \multirow{2}{*}{CdSe/ZnS} & Train & 3.380 & 1.311  & 1.057 & 3.231 \\
                                & Test  & 3.279 & 1.292  & 1.058 & 3.158 \\
    \end{tabular}
  \end{ruledtabular}
\end{table}

\begin{figure}[htbp]
  \centering
  \includegraphics[width=\singlefigure]{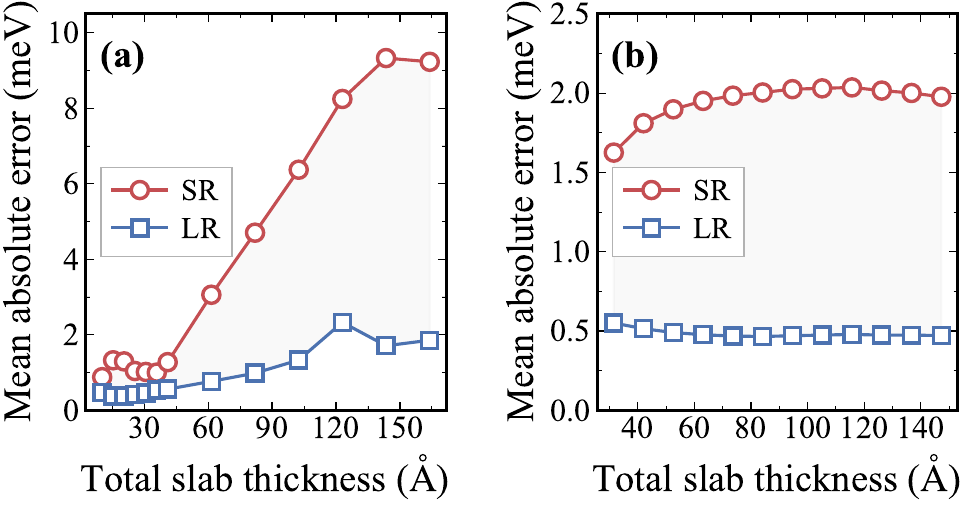}
  \caption{\label{fig2} Hamiltonian MAE of single-layer SR (red circles) and LR (blue squares) models versus layer thickness for (a)~GaN/AlN superlattices and (b)~polar ZnO slabs. In~(a), SR error grows with superlattice period and saturates at $\sim$140~\AA; in~(b), SR error rises with slab thickness before leveling off. The LR model [Eq.~(\ref{eq:longrange-ham})] maintains uniformly low errors across all thicknesses.}
\end{figure}

\begin{figure*}[t]
  \centering
  \includegraphics[width=0.75\textwidth]{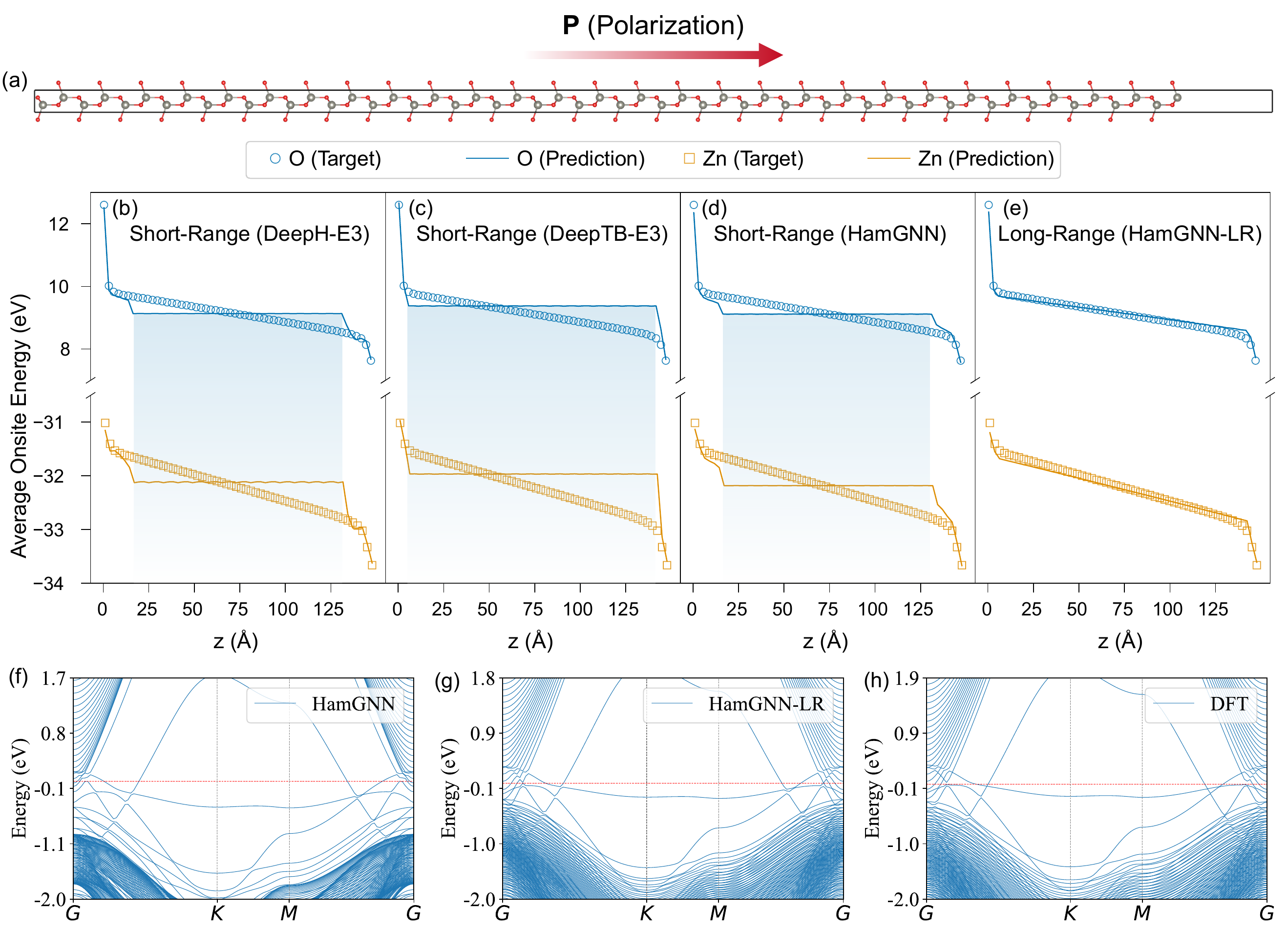}
  \caption{\label{fig3}
    Real-space onsite energy and band structure of the 56-layer polar ZnO slab.
    (a)~Structural schematic; red arrow denotes spontaneous polarization $\mathbf{P}$.
    (b)--(e)~Spatially resolved average onsite Hamiltonian matrix elements along $z$ for (b)~DeepH-E3, (c)~DeePTB-E3, (d)~short-range HamGNN, and (e)~HamGNN-LR. Open symbols: DFT reference; solid lines: model predictions. Shaded regions in (b)--(d) highlight staircase discontinuities arising from finite receptive fields; these artifacts are absent in (e) owing to the long-range correction.
    (f)--(h)~Band structures along $\Gamma$--$K$--$M$--$\Gamma$ for (f)~short-range HamGNN, (g)~HamGNN-LR, and (h)~DFT reference. Dashed red lines mark the Fermi level. HamGNN-LR closely reproduces DFT bands, whereas the short-range model exhibits visible deviations near the band gap.}
\end{figure*}

To examine size dependence, we evaluate single-layer SR and LR models as a function of layer thickness in GaN/AlN superlattices and polar ZnO slabs (Fig.~\ref{fig2}). In both systems, SR error rises with thickness: in GaN/AlN it grows with superlattice period and saturates at $\sim$140~\AA; in ZnO it increases with slab thickness before leveling off. The macroscopic built-in field, originating from bound polarization charges at interfaces (GaN/AlN) or spontaneous polarization along the $c$ axis (ZnO), produces an electrostatic potential that grows linearly beyond the single-layer receptive field. The LR model maintains uniformly low errors across all thicknesses, with only a mild increase at the largest sizes; even at maximum thickness, LR error remains below SR error at minimum thickness. This confirms that reciprocal-space Ewald correction in Eq.~(\ref{eq:longrange-ham}) successfully restores the macroscopic electrostatic potential discarded by real-space truncation.

We next benchmark HamGNN-LR against three representative short-range models---DeepH-E3~\cite{gong2023general}, DeePTB-E3~\cite{gu2024deeptb}, and HamGNN~\cite{zhong2023transferable,zhong2026unihamgnn}---all using three interaction layers. Beyond standard train/test evaluation on ZnO slabs, we perform a stringent size-extrapolation test: models trained on thinner slabs predict the Hamiltonian of a 56-layer ZnO slab far outside the training distribution (Table~\ref{tab:model_comparison}).

\begin{table}[htbp]
  \caption{\label{tab:model_comparison} Hamiltonian MAE (meV) for three-layer models on ZnO slabs. ``ZnO (thick)'' denotes size extrapolation to a 56-layer (147.25~\AA) slab unseen during training.}
  \footnotesize
  \begin{ruledtabular}
    \begin{tabular}{lcccc}
      Split       & DeepH-E3 & DeePTB-E3 & HamGNN & HamGNN-LR      \\
      \hline
      Train       & 0.462    & 1.940     & 0.397  & \textbf{0.291} \\
      Test        & 0.578    & 2.450     & 0.538  & \textbf{0.349} \\
      ZnO (thick) & 1.525    & 2.383     & 1.565  & \textbf{0.473} \\
    \end{tabular}
  \end{ruledtabular}
\end{table}

On the standard test set, HamGNN-LR achieves the lowest error (0.349~meV), a 35\% reduction over HamGNN (0.538~meV) and 40\% over DeepH-E3 (0.578~meV). This confirms that the long-range correction yields substantial gains even when the receptive field already spans three interaction layers. The advantage is most pronounced under size extrapolation: errors of all short-range models increase sharply---to 1.525~meV ($+$164\%) for DeepH-E3 and 1.565~meV ($+$191\%) for HamGNN---whereas HamGNN-LR reaches only 0.473~meV ($+$36\%). This robust extrapolation stems directly from Eq.~(\ref{eq:longrange-ham}): Ewald summation analytically captures the macroscopic spontaneous-polarization field along the $c$ axis, whose spatial extent grows linearly with slab thickness. No practically feasible number of message-passing layers can reproduce this behavior.

Figure~\ref{fig3} provides a microscopic view of the long-range correction on the 56-layer ZnO slab. In the DFT reference, onsite energies of both O and Zn atoms vary nearly linearly along $z$, reflecting the macroscopic electrostatic potential gradient generated by the spontaneous polarization. All three short-range models [Figs.~\ref{fig3}(b)--(d)] exhibit characteristic staircase artifacts: within each receptive-field window local chemistry is captured correctly, but the model cannot account for the electrostatic potential increment $\Delta\phi$ accumulated beyond the cutoff radius. The step width corresponds to the effective range of three message-passing layers, and the cumulative mismatch grows with $z$, consistent with the increasing errors reported in Table~\ref{tab:model_comparison}. HamGNN-LR [Fig.~\ref{fig3}(e)] eliminates these artifacts entirely. Ewald summation restores the continuous linear potential gradient by incorporating the small-$|\mathbf{k}|$ components of $S(\mathbf{k})$ that encode the macroscopic charge distribution. This advantage is mirrored in reciprocal space. The short-range band structure [Fig.~\ref{fig3}(f)] misplaces both valence- and conduction-band edges and distorts their curvature, yielding an underestimated band gap and spurious splittings of low-lying subbands along the high-symmetry path. In contrast, HamGNN-LR [Fig.~\ref{fig3}(g)] closely reproduces the DFT reference [Fig.~\ref{fig3}(h)]: band extrema, near-gap ordering, and curvature at $\Gamma$ and $M$ are all faithfully captured, confirming that long-range Ewald correction is essential for quantitative band-structure accuracy in polarized slabs.

\section{Conclusion}
We have derived closed-form, variationally consistent long-range Hamiltonian matrix elements [Eq.~(\ref{eq:longrange-ham})] in a nonorthogonal atomic-orbital basis and implemented them within HamGNN-LR, a framework that incorporates macroscopic Coulomb fields into Machine-learning electronic Hamiltonians. Single-layer ablation demonstrates that the explicit long-range correction is decisive in polar and heterostructure systems, whereas data-driven attention alone fails to reproduce long-range electrostatic effects. HamGNN-LR maintains low errors under size extrapolation to thick ZnO slabs outside the training distribution. Analysis of real-space onsite energy profiles traces the staircase artifacts of short-range models to truncated built-in fields and confirms that the long-range correction restores continuous potential gradients. These results demonstrate that analytical, variationally consistent long-range Coulomb corrections are essential for accurate, cross-scale Machine-learning electronic-structure predictions in polar materials and heterostructures. The framework opens a pathway toward accurate, large-scale electronic-structure modeling of interfaces, superlattices, and other systems in which long-range electrostatic effects are dominant.

\appendix
\bibliography{references}
\section{Derivation of the Long-Range Hamiltonian Correction}
\label{app:derivation}

This appendix provides a self-contained derivation of the
long-range Hamiltonian matrix elements
$H_{\text{lr},\mu\nu}$ stated in
Eq.~(\ref{eq:longrange-ham}).  The strategy is as follows.
The reciprocal-space electrostatic energy $E_{\text{lr}}$
depends on the density matrix $P_{\mu\nu}$ only through the
net atomic charges $\{Q_i\}$.  Applying the variational
principle $H_{\text{lr},\mu\nu}
= \partial E_{\text{lr}}/\partial P_{\mu\nu}$ therefore
reduces, via the chain rule, to evaluating two factors:
(i)~the \emph{density-matrix sensitivity}
$\partial Q_i/\partial P_{\mu\nu}$, which encodes how
atomic charges respond to changes in the density matrix, and
(ii)~the \emph{electrostatic response}
$\partial E_{\text{lr}}/\partial Q_i$, which gives the
long-range potential at each atomic site.  We derive each
factor in turn and then combine them into a closed-form
expression.

\subsection{Variational Framework and Chain Rule}
\label{app:framework}

The long-range electrostatic energy with Gaussian-smoothed
reciprocal-space cutoff is
[Eq.~(\ref{eq:longrange-energy})]
\begin{equation}
  E_{\text{lr}}
  = \frac{1}{2\epsilon_0 V}
    \sum_{0<|\mathbf{k}|<k_c}
    \frac{1}{k^2}\,
    e^{-\sigma^2 k^2/2}\,
    \bigl|S(\mathbf{k})\bigr|^2,
  \label{eq:Elr-app}
\end{equation}
where $V$ is the unit-cell volume, $\epsilon_0$ the vacuum
permittivity, $\sigma$ the Gaussian smoothing width, and the
structure factor
\begin{equation}
  S(\mathbf{k})
  = \sum_i Q_i\,e^{i\mathbf{k}\cdot\bm{\tau}_i}
  \label{eq:Sk-app}
\end{equation}
encodes the net atomic charges
$Q_i \equiv Z_i - q_i$ ($Z_i$: nuclear charge;
$q_i$: electronic population) at positions $\bm{\tau}_i$.
We assume overall charge neutrality,
$\sum_i Q_i = 0$, a prerequisite for the convergence of the
Ewald sum in periodic systems and for the well-definedness of
$S(\mathbf{k})$ at small $|\mathbf{k}|$.

Because $E_{\text{lr}}$ depends on $P_{\mu\nu}$ only through
$\{Q_i\}$, the variational principle yields
\begin{equation}
  \begin{split}
    H_{\text{lr},\mu\nu}
    &= \frac{\partial E_{\text{lr}}}{\partial P_{\mu\nu}} \\[6pt]
    &= \sum_i
      \underbrace{%
        \frac{\partial E_{\text{lr}}}{\partial Q_i}
      }_{\text{electrostatic response}}
      \;\;
      \underbrace{%
        \frac{\partial Q_i}{\partial P_{\mu\nu}}
      }_{\text{density-matrix sensitivity}}.
  \end{split}
  \label{eq:chain-app}
\end{equation}
This equation serves as the roadmap for the remainder of the
derivation.  We evaluate the right-hand factor first
(Sec.~\ref{app:dQdP}), then the left-hand factor
(Sec.~\ref{app:dEdQ}), and finally assemble the result
(Sec.~\ref{app:assembly}).

\subsection{Density-Matrix Sensitivity:
  \texorpdfstring{$\partial Q_i/\partial P_{\mu\nu}$}{dQ/dP}}
\label{app:dQdP}

In a nonorthogonal atomic-orbital (NAO) basis
$\{\phi_\mu(\mathbf{r})\}$, the single-particle density
matrix $P_{\mu\nu}$ determines the electron density
\begin{equation}
  \rho(\mathbf{r})
  = \sum_{\mu\nu} P_{\mu\nu}\,
    \phi_\mu^*(\mathbf{r})\,\phi_\nu(\mathbf{r}).
  \label{eq:rho-app}
\end{equation}
The electronic population on atom~$i$ is obtained by
projecting $\rho(\mathbf{r})$ onto a spatially localized
weight function
$W_i(\mathbf{r}-\bm{\tau}_i)$~\cite{hirshfeld1977,becke1988multicenter}:
\begin{equation}
  q_i
  = \int W_i(\mathbf{r}-\bm{\tau}_i)\,
    \rho(\mathbf{r})\,\mathrm{d}\mathbf{r}
  = \sum_{\mu\nu} P_{\mu\nu}\,W_{i,\mu\nu},
  \label{eq:qi-app}
\end{equation}
where the second equality follows from substituting
Eq.~(\ref{eq:rho-app}) and defining the orbital-projected
weight-matrix elements
\begin{equation}
  W_{i,\mu\nu}
  \equiv \int W_i(\mathbf{r}-\bm{\tau}_i)\,
         \phi_\mu^*(\mathbf{r})\,\phi_\nu(\mathbf{r})\,
         \mathrm{d}\mathbf{r}.
  \label{eq:Wimu-app}
\end{equation}
Because $Q_i = Z_i - q_i$ and $Z_i$ is independent of
$P_{\mu\nu}$, the desired sensitivity is
\begin{equation}
  \frac{\partial Q_i}{\partial P_{\mu\nu}}
  = -\frac{\partial q_i}{\partial P_{\mu\nu}}
  = -W_{i,\mu\nu}.
  \label{eq:dQdP-app}
\end{equation}
For a real-valued NAO basis---which can always be chosen when
time-reversal symmetry holds---$W_{i,\mu\nu}$ is real,
a property that will be used when simplifying the final
expression.

\subsection{Electrostatic Response:
  \texorpdfstring{$\partial E_{\text{lr}}/\partial Q_i$}{dE/dQ}}
\label{app:dEdQ}

To evaluate the second factor in
Eq.~(\ref{eq:chain-app}), we differentiate
$|S(\mathbf{k})|^2 = S(\mathbf{k})\,S^*(\mathbf{k})$
with respect to~$Q_i$.  The product rule gives
\begin{equation}
  \frac{\partial|S|^2}{\partial Q_i}
  = \frac{\partial S}{\partial Q_i}\,S^*
  + S\,\frac{\partial S^*}{\partial Q_i}.
  \label{eq:dS2dQ-product}
\end{equation}
From Eq.~(\ref{eq:Sk-app}),
$\partial S/\partial Q_i = e^{i\mathbf{k}\cdot\bm{\tau}_i}$
and
$\partial S^*/\partial Q_i = e^{-i\mathbf{k}\cdot\bm{\tau}_i}$.
Substituting and using $z + z^* = 2\operatorname{Re}(z)$
with $z \equiv S(\mathbf{k})\,e^{-i\mathbf{k}\cdot\bm{\tau}_i}$:
\begin{equation}
  \frac{\partial|S(\mathbf{k})|^2}{\partial Q_i}
  = 2\,\operatorname{Re}\!\bigl[
      S(\mathbf{k})\,e^{-i\mathbf{k}\cdot\bm{\tau}_i}
    \bigr].
  \label{eq:dS2-result}
\end{equation}
Differentiating Eq.~(\ref{eq:Elr-app}) and noting that the
factor $\tfrac{1}{2}$ in $E_{\text{lr}}$ cancels with the
factor of $2$ from Eq.~(\ref{eq:dS2-result}), we obtain
\begin{equation}
  \frac{\partial E_{\text{lr}}}{\partial Q_i}
  = \frac{1}{\epsilon_0 V}
    \sum_{0<|\mathbf{k}|<k_c}
    \frac{e^{-\sigma^2 k^2/2}}{k^2}\,
    \operatorname{Re}\!\bigl[
      S(\mathbf{k})\,e^{-i\mathbf{k}\cdot\bm{\tau}_i}
    \bigr].
  \label{eq:dEdQ-app}
\end{equation}
Physically, this quantity equals the long-range component of
the electrostatic potential at site~$i$, generated by the
Gaussian-smoothed charge distribution of all other atoms in
reciprocal space.

\subsection{Assembly and Final Result}
\label{app:assembly}

We now substitute
Eqs.~(\ref{eq:dQdP-app}) and~(\ref{eq:dEdQ-app}) into the
chain rule [Eq.~(\ref{eq:chain-app})]:
\begin{equation}
  \begin{split}
    H_{\text{lr},\mu\nu}
    &= -\frac{1}{\epsilon_0 V}
      \sum_{0<|\mathbf{k}|<k_c}
      \frac{e^{-\sigma^2 k^2/2}}{k^2} \\
    &\quad\times
      \operatorname{Re}\!\Bigl[
        S(\mathbf{k})
        \sum_i e^{-i\mathbf{k}\cdot\bm{\tau}_i}
        W_{i,\mu\nu}
      \Bigr].
  \end{split}
  \label{eq:Hlr-step1}
\end{equation}
Because $W_{i,\mu\nu}$ is real
(Sec.~\ref{app:dQdP}), it can be moved inside the
$\operatorname{Re}[\cdots]$ operator, and the sum over
atoms $i$ can be performed under the same sign:
\begin{equation}
  \begin{split}
    H_{\text{lr},\mu\nu}
    = -\frac{1}{\epsilon_0 V}
      \sum_{0<|\mathbf{k}|<k_c}
      \frac{e^{-\sigma^2 k^2/2}}{k^2}
    \\
    \times\operatorname{Re}\!\Bigl[
      S(\mathbf{k})
      \sum_i e^{-i\mathbf{k}\cdot\bm{\tau}_i}\,
      W_{i,\mu\nu}
    \Bigr].
  \end{split}
  \label{eq:Hlr-SI}
\end{equation}
Converting to Hartree atomic units
($4\pi\epsilon_0 = 1$, hence $1/\epsilon_0 = 4\pi$) yields
the final closed-form expression:
\begin{equation}
  \begin{split}
    H_{\text{lr},\mu\nu}
    = -\frac{4\pi}{V}
      \sum_{0<|\mathbf{k}|<k_c}
      \frac{e^{-\sigma^2 k^2/2}}{k^2}
    \\
    \times\operatorname{Re}\!\Bigl[
      S(\mathbf{k})
      \sum_i e^{-i\mathbf{k}\cdot\bm{\tau}_i}\,
      W_{i,\mu\nu}
    \Bigr],
  \end{split}
  \label{eq:Hlr-final-app}
\end{equation}
which is Eq.~(\ref{eq:longrange-ham}) of the main text.
By construction,
$H_{\text{lr},\mu\nu}
= \partial E_{\text{lr}}/\partial P_{\mu\nu}$, ensuring
variational consistency between the long-range energy and
Hamiltonian.

\paragraph{Pairwise phase structure.}
Expanding $S(\mathbf{k}) = \sum_j Q_j\,
e^{i\mathbf{k}\cdot\bm{\tau}_j}$ in
Eq.~(\ref{eq:Hlr-final-app}), the argument of
$\operatorname{Re}[\cdots]$ becomes
\begin{equation}
  \sum_{i,j}
    Q_j\,W_{i,\mu\nu}\,
    e^{-i\mathbf{k}\cdot(\bm{\tau}_i-\bm{\tau}_j)},
  \label{eq:pairwise-app}
\end{equation}
a sum over atom pairs $(i,j)$ whose phase depends solely on
the relative displacement
$\bm{\tau}_i - \bm{\tau}_j$.  This translational-invariant
pairwise structure is the key property exploited by the
Ewald attention module
(Appendix~\ref{app:ewald_attention}).

\subsection{Ewald Self-Interaction Correction}
\label{app:self}

The reciprocal-space Ewald sum in
Eq.~(\ref{eq:Elr-app}) implicitly includes the unphysical
interaction of each point charge with its own Gaussian
screening cloud~\cite{toukmaji1996ewald,deleeuw1980ewald}.
This self-interaction must be subtracted.  We derive the
corresponding Hamiltonian correction by the same variational
chain-rule procedure used above.

\paragraph{Self-interaction energy.}
The Gaussian screening cloud centered on atom~$i$ has the
normalized charge density
\begin{equation}
  \rho_i^{\text{scr}}(\mathbf{r})
  = \frac{Q_i}{(2\pi\sigma^2)^{3/2}}\,
    \exp\!\Bigl(
      -\frac{|\mathbf{r}-\bm{\tau}_i|^2}{2\sigma^2}
    \Bigr).
  \label{eq:rho-scr}
\end{equation}
The electrostatic potential of this distribution, evaluated
at its own center $\mathbf{r}=\bm{\tau}_i$, is
\begin{equation}
  \phi_i^{\text{self}}
  = \frac{1}{4\pi\epsilon_0}
    \int
    \frac{\rho_i^{\text{scr}}(\mathbf{r})}
         {|\mathbf{r}-\bm{\tau}_i|}\,
    \mathrm{d}^3\mathbf{r}
  = \frac{Q_i}{4\pi\epsilon_0}\,
    \sqrt{\frac{2}{\pi}}\,\frac{1}{\sigma},
  \label{eq:phi-self}
\end{equation}
where the radial integral
$\int_0^\infty r\,e^{-r^2/(2\sigma^2)}\,\mathrm{d}r
= \sigma^2$ has been used.  The total self-interaction
energy, with the factor of $\tfrac{1}{2}$ to avoid double
counting, is
\begin{equation}
  E_{\text{self}}
  = \frac{1}{2}\sum_i Q_i\,\phi_i^{\text{self}}
  = \frac{1}{4\pi\epsilon_0}\,
    \frac{1}{\sqrt{2\pi}\,\sigma}
    \sum_i Q_i^2.
  \label{eq:E-self}
\end{equation}

\paragraph{Variational Hamiltonian correction.}
Differentiating Eq.~(\ref{eq:E-self}) with respect to
$Q_i$ gives
$\partial E_{\text{self}}/\partial Q_i
= 2Q_i/(4\pi\epsilon_0\sqrt{2\pi}\,\sigma)$.
Applying the chain rule
$H_{\text{self},\mu\nu}
= \sum_i
  (\partial E_{\text{self}}/\partial Q_i)\,
  (\partial Q_i/\partial P_{\mu\nu})$
with $\partial Q_i/\partial P_{\mu\nu} = -W_{i,\mu\nu}$
[Eq.~(\ref{eq:dQdP-app})] yields, after conversion to
Hartree atomic units ($4\pi\epsilon_0=1$),
\begin{equation}
  H_{\text{self},\mu\nu}
  = -\frac{2}{\sqrt{2\pi}\,\sigma}
    \sum_i Q_i\,W_{i,\mu\nu}.
  \label{eq:Hself-final}
\end{equation}

\paragraph{Locality and absorption into $H_{\text{sr}}$.}
In contrast to the reciprocal-space sum in
$H_{\text{lr},\mu\nu}$
[Eq.~(\ref{eq:Hlr-final-app})], the self-interaction
correction Eq.~(\ref{eq:Hself-final}) is
\emph{spatially local}: $W_{i,\mu\nu}$ is nonvanishing only
for orbitals $\mu,\nu$ within the compact support of the
weight function centered on atom~$i$, so
$H_{\text{self},\mu\nu}$ contributes only to onsite and
near-neighbor matrix elements.  The complete variationally
consistent long-range Hamiltonian is
\begin{equation}
  H_{\mu\nu}^{\text{lr,full}}
  = H_{\text{lr},\mu\nu}
  - H_{\text{self},\mu\nu}.
  \label{eq:Hlr-full}
\end{equation}
Because $H_{\text{self},\mu\nu}$ lies entirely within the
receptive field of the short-range message-passing network,
it can equivalently be absorbed into the short-range channel
$H_{\text{sr}}$ during training.  In either case, variational
consistency is maintained:
$H_{\alpha,\mu\nu}
= \partial E_\alpha/\partial P_{\mu\nu}$ holds for each
component $\alpha\in\{\text{sr},\,\text{lr},\,\text{self}\}$
independently, so
\begin{equation}
  H_{\text{tot},\mu\nu}
  = H_{\text{sr},\mu\nu}
  + H_{\text{lr},\mu\nu}
  - H_{\text{self},\mu\nu}
  = \frac{\partial E_{\text{tot}}}{\partial P_{\mu\nu}}
\end{equation}
is guaranteed by construction.

\section{Algorithmic Details of the Ewald Attention Module}
\label{app:ewald_attention}

This appendix specifies the Ewald attention module
introduced in the main text.  The module takes
E(3)-equivariant node features
$\{\mathbf{h}_i\}_{i=1}^{N}$ produced by short-range
message passing, extracts long-range correlations in
reciprocal space, and outputs
(i)~updated long-range node features
$\mathbf{h}_i^{\text{lr}}$,
(ii)~effective ionic charges $Q_i$, and
(iii)~orbital-projected weight matrices $W_{i,\mu\nu}$,
which are substituted into
Eq.~(\ref{eq:longrange-ham}) to yield $H_{\text{lr}}$.
The computation proceeds in five stages:
reciprocal-grid setup
(Sec.~\ref{app:grid}),
equivariant feature projection
(Sec.~\ref{app:proj}),
rotary position encoding
(Sec.~\ref{app:rope}),
linearized aggregation and feature update
(Sec.~\ref{app:agg}), and
decoding and $H_{\text{lr}}$ assembly
(Sec.~\ref{app:decode}).

Throughout, we adopt the following notation: $N$ is the
number of atoms in the periodic cell, $\bm{\tau}_j$ the
position of atom~$j$,
$\mathbf{C}\in\mathbb{R}^{3\times 3}$ the real-space
lattice matrix, and $d$ the feature dimension.

\subsection{Reciprocal Grid and Base Weights}
\label{app:grid}

The reciprocal lattice vectors are
$\mathbf{B} = 2\pi(\mathbf{C}^{-1})^\top$.  We enumerate
all reciprocal-lattice points satisfying
$0 < |\mathbf{k}_m| < k_c$ to form the discrete wave-vector
set $\mathcal{K}=\{\mathbf{k}_m\}_{m=1}^{M}$, where $k_c$
is determined by an energy cutoff $E_{\text{cut}}$.

Each wave vector carries a scalar base weight
$w(\mathbf{k}_m)$ that modulates its contribution to the
attention.  We support two choices:
\begin{equation}
  w(\mathbf{k}_m) =
  \begin{cases}
    \displaystyle
    \frac{1}{|\mathbf{k}_m|^2}\,
    \exp\!\Bigl(-\frac{\sigma^2|\mathbf{k}_m|^2}{2}\Bigr),
    & \text{(Ewald)}
    \\[8pt]
    \text{Softplus}\!\bigl(
      \text{MLP}(|\mathbf{k}_m|^2)
    \bigr),
    & \text{(adaptive)}
  \end{cases}
  \label{eq:base-weight}
\end{equation}
where $\sigma$ is the Gaussian smoothing width.  The Ewald
form reproduces the standard reciprocal-space kernel; the
adaptive form allows the network to learn
$|\mathbf{k}|$-dependent modulations while remaining
positive by construction.

\subsection{Equivariant Feature Projection}
\label{app:proj}

A central design requirement is that the attention module
preserves E(3) equivariance end to end.  We achieve this by
separating the roles of queries/keys (which determine
\emph{how much} information flows) from values (which carry
\emph{what} information flows):

\begin{itemize}
\item \textbf{Queries and keys} are derived from the
  \emph{invariant} ($l=0$) scalar component
  $\mathbf{x}_j^{(0)}$ of the equivariant node feature
  $\mathbf{h}_j$:
  \begin{equation}
    \mathbf{q}_j
    = \operatorname{SiLU}\!\bigl(
        \mathbf{x}_j^{(0)}\,\mathbf{W}_Q
      \bigr),
    \quad
    \mathbf{k}_j
    = \operatorname{SiLU}\!\bigl(
        \mathbf{x}_j^{(0)}\,\mathbf{W}_K
      \bigr),
    \label{eq:qk-proj}
  \end{equation}
  with learnable matrices
  $\mathbf{W}_Q,\mathbf{W}_K
    \in\mathbb{R}^{d_0\times d}$ and SiLU activation for
  improved expressiveness and numerical stability under
  subsequent rotary encoding.
  Because $\mathbf{q}_j$ and $\mathbf{k}_j$ depend only on
  scalars, they are invariant under rotation and can be freely
  combined with position-dependent phases (Sec.~\ref{app:rope})
  without breaking equivariance.

\item \textbf{Values} are taken directly from the full
  equivariant feature:
  $\mathbf{v}_j = \mathbf{h}_j$.
  No position encoding is applied to values, so their
  equivariant transformation properties are preserved
  throughout aggregation.
\end{itemize}

\noindent
This separation ensures that attention weights are
rotationally invariant scalars, while the aggregated output
inherits equivariance from the value channel alone.

\subsection{Rotary Position Encoding and Ewald Phase
  Recovery}
\label{app:rope}

To inject spatial information into the attention weights, we
apply rotary position encoding
(RoPE)~\cite{su2024roformer} to queries and keys, extending
RoPE from discrete sequence positions to continuous
real-space phases defined by the reciprocal lattice.

\paragraph{Definition.}
For each atom~$j$ and wave vector $\mathbf{k}_m$, we define
the scalar phase
$\theta_{j,m} = \mathbf{k}_m\cdot\bm{\tau}_j$.
The feature vector of even dimension~$d$ is partitioned into
$d/2$ two-dimensional blocks, each rotated by $\theta_{j,m}$.
Concretely, the block-diagonal rotation matrix
$\mathbf{R}^{d}(\theta)
  = \operatorname{diag}\!\bigl(
    \mathbf{R}(\theta),\dots,\mathbf{R}(\theta)
  \bigr)$
with $2\times 2$ blocks
\begin{equation}
  \mathbf{R}(\theta)
  = \begin{pmatrix}
      \cos\theta & -\sin\theta \\
      \sin\theta & \;\;\cos\theta
    \end{pmatrix}
  \label{eq:R2d}
\end{equation}
is applied to produce encoded queries and keys:
\begin{equation}
  \tilde{\mathbf{q}}_{j,m}
  = \mathbf{R}^{d}(\theta_{j,m})\,\mathbf{q}_j,
  \qquad
  \tilde{\mathbf{k}}_{j,m}
  = \mathbf{R}^{d}(\theta_{j,m})\,\mathbf{k}_j.
  \label{eq:rope-encode}
\end{equation}
All $d/2$ blocks share the same physical phase
$\theta_{j,m}$, because
$\mathbf{k}_m\cdot\bm{\tau}_j$ is a single scalar.
Equivalently, each 2D block can be viewed as multiplying the
complex number $q_{j,1}+i\,q_{j,2}$ by $e^{i\theta_{j,m}}$.

\paragraph{Sign convention.}
Both queries and keys use the \emph{same} sign for the phase
angle.  This choice is essential: if one were to negate the
phase for keys (as in some RoFormer variants), the resulting
inner product would depend on
$\mathbf{k}_m\cdot(\bm{\tau}_i+\bm{\tau}_j)$, which is
\emph{not} translationally invariant and does not match the
Ewald pairwise structure.

\paragraph{Recovery of the Ewald relative phase.}
The inner product of encoded query and key evaluates to
\begin{equation}
  \bigl\langle
    \tilde{\mathbf{q}}_{i,m},\;
    \tilde{\mathbf{k}}_{j,m}
  \bigr\rangle
  = \mathbf{q}_i^\top\,
    \mathbf{R}^{d}(\theta_{j,m}-\theta_{i,m})\,
    \mathbf{k}_j,
  \label{eq:qk-inner}
\end{equation}
where we used the orthogonality
$\mathbf{R}^{d}(\theta)^\top
  = \mathbf{R}^{d}(-\theta)$.
The effective rotation angle is
\begin{equation}
  \theta_{j,m}-\theta_{i,m}
  = \mathbf{k}_m\cdot(\bm{\tau}_j-\bm{\tau}_i)
  = -\mathbf{k}_m\cdot(\bm{\tau}_i-\bm{\tau}_j),
\end{equation}
so the inner product reduces to a linear combination of
$\cos\!\bigl[\mathbf{k}_m\cdot(\bm{\tau}_i-\bm{\tau}_j)\bigr]$
and
$\sin\!\bigl[\mathbf{k}_m\cdot(\bm{\tau}_i-\bm{\tau}_j)\bigr]$,
with coefficients determined by components of
$\mathbf{q}_i$ and $\mathbf{k}_j$ in each 2D block.
This is precisely the relative-displacement phase
$\mathbf{k}_m\cdot(\bm{\tau}_i-\bm{\tau}_j)$ that appears
in the pairwise expansion of
Eq.~(\ref{eq:longrange-ham})
[cf.\ Eq.~(\ref{eq:pairwise-app})], confirming that the
RoPE-based attention kernel reproduces the Ewald summation
phase structure while respecting translational invariance.

\subsection{Linearized Aggregation and Feature Update}
\label{app:agg}

\paragraph{Reciprocal-space context matrix.}
Standard all-pairs attention scales as
$\mathcal{O}(N^2)$.  We instead adopt linearized attention,
factorizing the computation as
$\mathbf{Q}(\mathbf{K}^\top\mathbf{V})$, which reduces the
cost to $\mathcal{O}(N)$ per wave vector and
$\mathcal{O}(N|\mathcal{K}|)$ overall.

For each $\mathbf{k}_m$, we first aggregate all atoms into a
global context matrix
$\mathbf{C}_m\in\mathbb{R}^{d\times d_v}$:
\begin{equation}
  \mathbf{C}_m
  = \frac{w(\mathbf{k}_m)}{N}
    \sum_{j=1}^{N}
    \tilde{\mathbf{k}}_{j,m}\otimes\mathbf{v}_j,
  \label{eq:context}
\end{equation}
where $w(\mathbf{k}_m)$ is the base weight
[Eq.~(\ref{eq:base-weight})] and the outer product
$\tilde{\mathbf{k}}_{j,m}\otimes\mathbf{v}_j$ contracts
over the key dimension while retaining equivariant structure
in the value dimension.

\paragraph{Per-node feature extraction.}
Each atom~$i$ retrieves its long-range feature by
contracting its query with the context matrices and summing
over wave vectors:
\begin{equation}
  \mathbf{h}_i^{\text{lr}}
  = \frac{1}{\sqrt{d}}
    \sum_{\mathbf{k}_m\in\mathcal{K}}
    \mathbf{C}_m^\top\,\tilde{\mathbf{q}}_{i,m},
  \label{eq:hlr}
\end{equation}
with the standard $1/\sqrt{d}$ scaling.  Because the
contraction $\mathbf{C}_m^\top\tilde{\mathbf{q}}_{i,m}$
involves an invariant query and the equivariant value
content stored in $\mathbf{C}_m$, the output
$\mathbf{h}_i^{\text{lr}}$ inherits E(3) equivariance from
the value channel.

Note that the attention weight in Eq.~(\ref{eq:context})
uses only the base weight $w(\mathbf{k}_m)$ and does not
depend on the charge-dependent structure factor
$S(\mathbf{k}_m)$.  The module thus serves as a
\emph{feature extractor}: it produces long-range-aware node
representations from which charges and weight matrices are
decoded in a subsequent step (Sec.~\ref{app:decode}).

\paragraph{Gated residual connection.}
The short-range and long-range channels are fused via a
gated residual:
\begin{equation}
  \mathbf{h}_i^{\text{lr}}
  \leftarrow
  (1-\alpha)\,\mathbf{h}_i^{\text{sr}}
  + \alpha\;\mathrm{EquivariantNonlin}
    (\mathbf{h}_i^{\text{lr}}),
  \label{eq:gate}
\end{equation}
where $\alpha\in[0,1]$ is a learnable scalar gate
initialized near zero.  This design mirrors the energy
decomposition
$E_{\text{tot}}=E_{\text{sr}}+E_{\text{lr}}$ at the
feature level: the short-range baseline is preserved intact
while the long-range correction enters perturbatively,
ensuring stable early training and automatic suppression in
nonpolar systems.  Because the nonlinearity acts
channel-wise without mixing angular-momentum components,
$\mathbf{h}_i^{\text{lr}}$ retains E(3) equivariance.

\subsection{Decoding and $H_{\text{lr}}$ Assembly}
\label{app:decode}

The updated node features $\mathbf{h}_i^{\text{lr}}$ are
decoded into all quantities required by
Eq.~(\ref{eq:longrange-ham}).

\paragraph{Effective ionic charges.}
A scalar charge $Q_i\in\mathbb{R}$ is obtained by
extracting the $l=0$ component of
$\mathbf{h}_i^{\text{lr}}$ followed by a linear layer.
Global neutrality is enforced by a zero-mean projection:
\begin{equation}
  \tilde{Q}_j
  = Q_j - \frac{1}{N}\sum_{k=1}^{N}Q_k.
  \label{eq:neutralize}
\end{equation}

\paragraph{Long-range edge features.}
For each pair $(i,j)$ within the real-space neighbor cutoff,
we construct edge features by concatenating the node
features of both endpoints and coupling them to the
spherical-harmonic embedding of the bond direction
$\hat{\mathbf{r}}_{ij}
  =(\bm{\tau}_j-\bm{\tau}_i)
   /|\bm{\tau}_j-\bm{\tau}_i|$
via a Clebsch--Gordan tensor product:
\begin{equation}
  \mathbf{e}_{ij}^{\text{lr}}
  = \bigl(
      \mathbf{h}_i^{\text{lr}}
      \oplus
      \mathbf{h}_j^{\text{lr}}
    \bigr)
    \otimes Y^{(l)}(\hat{\mathbf{r}}_{ij}),
  \label{eq:edge-lr}
\end{equation}
where $\oplus$ denotes concatenation and $\otimes$ the
Clebsch--Gordan product coupling the irreducible
representations of the node features with those of the
spherical harmonics $Y^{(l)}$.  Under rotation, both
inputs transform as SO(3) irreps, and the coupling ensures
that $\mathbf{e}_{ij}^{\text{lr}}$ transforms accordingly,
preserving equivariance.

\paragraph{Weight matrices.}
The orbital-projected weight matrices $W_{i,\mu\nu}$ are
decoded from node features (onsite elements with both
$\mu,\nu$ on atom~$i$) and edge features (offsite elements
with $\mu$ on $i$ and $\nu$ on neighbor~$j$).  Both
decoders respect equivariance constraints imposed by the
angular-momentum quantum numbers of the orbitals.

\paragraph{Structure factor and $H_{\text{lr}}$.}
With $\tilde{Q}_j$ and $W_{i,\mu\nu}$ in hand, we compute
the structure factor at each sampled wave vector,
\begin{equation}
  S(\mathbf{k}_m)
  = \frac{1}{N}\sum_{j=1}^{N}\tilde{Q}_j\,
    \exp(i\,\mathbf{k}_m\cdot\bm{\tau}_j),
  \label{eq:Sk-module}
\end{equation}
where the $1/N$ prefactor renders $S(\mathbf{k}_m)$ an
intensive quantity consistent with the $1/V$ normalization
in Eq.~(\ref{eq:longrange-energy}).  The long-range
Hamiltonian correction follows by direct substitution into
Eq.~(\ref{eq:longrange-ham}):
\begin{equation}
  \begin{split}
    H_{\text{lr},\mu\nu}
    = -\frac{4\pi}{V}
      \sum_{0<|\mathbf{k}_m|<k_c}
      \frac{e^{-\sigma^2 k_m^2/2}}{k_m^2}
    \\
    \times\operatorname{Re}\!\Bigl[
      S(\mathbf{k}_m)\sum_i
      e^{-i\mathbf{k}_m\cdot\bm{\tau}_i}\,
      W_{i,\mu\nu}
    \Bigr],
  \end{split}
  \label{eq:Hlr-module}
\end{equation}
evaluated analytically once $\tilde{Q}_j$ and
$W_{i,\mu\nu}$ are provided by the network.  The total
predicted Hamiltonian,
\begin{equation}
  H_{\mu\nu}
  = H_{\text{sr},\mu\nu}
  + H_{\text{lr},\mu\nu},
\end{equation}
completes the dual-channel pipeline.
\end{document}